\documentclass[journal,12pt,draftcls,onecolumn,twoside]{IEEEtran}

\usepackage[cmex10]{amsmath}
\usepackage{amssymb}
\usepackage{epsfig}
\usepackage{subfigure}
\usepackage{graphicx}

\def\BibTeX{{\rm B\kern-.05em{\sc i\kern-.025em b}\kern-.08em
    T\kern-.1667em\lower.7ex\hbox{E}\kern-.125emX}}

\title{Downlink Interference Alignment}

\author{Changho Suh, Minnie Ho$^\dag$ and David Tse\\

Wireless Foundations \\
University of California at Berkeley \\
Email: \{chsuh, dtse\}@eecs.berkeley.edu

$^\dag$Intel Labs, Intel Corporation\\
Email: minnie.ho@intel.com
\thanks{This work was supported by a gift from Intel and a grant CNS-0722032 from the National Science Foundation.}}

\begin{document}

\IEEEoverridecommandlockouts

\maketitle

\begin{abstract}
We develop an interference alignment (IA) technique for a downlink cellular system. In the uplink, IA schemes need channel-state-information exchange \emph{across base-stations of different cells}, but our downlink IA technique requires feedback \emph{only within a cell}. As a result, the proposed scheme can be implemented with a few changes to an existing cellular system where the feedback mechanism (within a cell) is already being considered for supporting multi-user MIMO. Not only is our proposed scheme implementable with little effort, it can in fact provide substantial gain especially when interference from a dominant interferer is significantly stronger than the remaining interference: it is shown that in the two-isolated cell layout, our scheme provides four-fold gain in throughput performance over a standard multi-user MIMO technique. We show through simulations that our technique provides respectable gain under a more realistic scenario: it gives approximately 20\% gain for a 19 hexagonal wrap-around-cell layout. Furthermore, we show that our scheme has the potential to provide substantial gain for \emph{macro-pico} cellular networks where pico-users can be significantly interfered with by the nearby macro-BS.
\end{abstract}

\begin{keywords}
Interference Alignment, Downlink, Multi-User MIMO, Macro-Pico Cellular Networks
\end{keywords}

\section{Introduction}

One of the key performance metrics in the design of cellular systems is that of cell-edge spectral efficiency.
As a result, fourth-generation (4G) cellular systems, such as WiMAX \cite{80216m} and 3GPP-LTE \cite{3GPPLTE}, require at least a doubling in cell-edge throughput over previous 3G systems \cite{3GPPLTE}. Given the disparity between average and cell-edge spectral efficiencies (ratios of about 4:1) \cite{80216m}, the desire to improve cell-edge throughput performance is likely to continue.

Since the throughput of cell-edge users is greatly limited by the presence of co-channel interference from other cells, developing an intelligent interference management scheme is the key to improving cell-edge throughput. One interesting recent development, called \emph{interference alignment} (IA) \cite{Mohammad, Jafar:IC}, manages interference by aligning multiple interference signals in a signal subspace with dimension smaller than the number of interferers. While most of the work on IA \cite{Jafar:IC,Jafar:DB,Heath:First} has focused on $K$ point-to-point interfering links, it has also been shown in \cite{SuhTse:IA,Caire:IA,Tresch:CellularIA} that IA can be used to improve the cell-edge user throughput in a cellular network. Especially, it was shown in \cite{SuhTse:IA} that \emph{near interference-free throughput} performance can be achieved in the cellular network.

While IA promises substantial theoretical gain in cellular networks, it comes with challenges in implementation. First, the uplink IA scheme in \cite{SuhTse:IA} requires extensive channel-state-information (CSI) to be exchanged over the backhaul \emph{between base-stations (BSs) of different cells}.
A second challenge comes from realistic cellular environments that involve multiple unaligned out-of-cell interferers.
Lastly, the integration of IA with other system issues, such as scheduling, needs to be addressed.

We propose a new IA technique for downlink cellular systems that addresses many of these practical concerns. Unlike the uplink IA, our downlink IA scheme requires feedback only within a cell. As a consequence, our technique can be implemented with small changes to existing 4G standards where the within-a-cell feedback mechanism is already being considered for supporting multi-user MIMO.
Our proposed technique builds on the idea of the IA technique in \cite{SuhTse:IA} that aims for a two-isolated cell layout and can thus cancel interference only from one neighboring BS.
We observe that the IA technique in \cite{SuhTse:IA} may give up the opportunity of providing matched-filtered gain (also called beam-forming gain in the case of multiple antennas) in the presence of a large number of interferers. Our new technique balances these two scenarios, inspired by the idea of the standard MMSE receiver that unifies a zero-forcing receiver (optimum in the high $\sf SNR$ regime) and a matched filter (optimum in the low $\sf SNR$ regime).

Through simulations, we show that our scheme provides approximately 55\% and 20\% gain in cell-edge throughput performance for a linear cell layout and 19 hexagonal wrap-around-cell layout, respectively, as compared to a standard multi-user MIMO technique. We also find that our scheme has the potential to provide significant performance for heterogeneous networks~\cite{Heterogeneous}, e.g., macro-pico cellular networks where a dominant interference can be much stronger than the residual interference. For instance, pico-users can be significantly interfered with by the nearby macro-BS, as compared to the aggregated remaining BSs. We show that for these networks our scheme can give around 30\% to 200\% gain over the standard technique. Furthermore, our scheme is easily combined with a widely-employed opportunistic scheduler~\cite{Tse:Book} for significant multi-user-diversity gain.


\section{Interference Alignment}
\label{sec-IAtechnique}

\subsection{Review of Uplink IA}

\begin{figure}[t]
\begin{center}
{\epsfig{figure=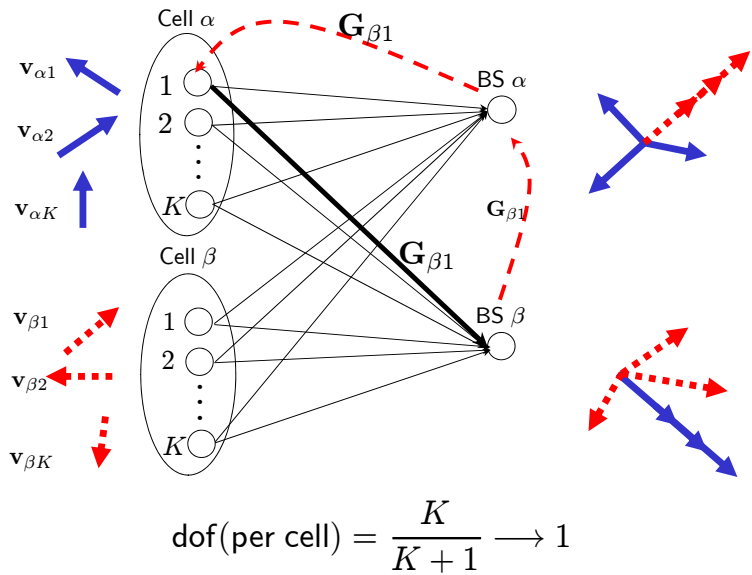, angle=0, width=0.6\textwidth}}
\end{center}
\caption{Uplink interference alignment. Interference-free degrees-of-freedom can be asymptotically achieved with an increase in $K$. However, this scheme requires exchange of cross-channel information over the backhaul between BSs of different cells.
}
\label{fig:uplinkIA}
\end{figure}

We begin by reviewing uplink IA in~\cite{SuhTse:IA}. Fig. \ref{fig:uplinkIA} illustrates an example for the case of two isolated cells $\alpha$ and $\beta$. Suppose that each cell has $K$ users and each user (e.g., user $k$ in cell $\alpha$) sends one symbol (or stream) along a transmitted vector $\mathbf{v}_{\alpha k} \in \mathbb{C}^{M}$.
Each user can generate multiple dimensions by using subcarriers (in an OFDM system), antennas, or both:
\begin{align}
M = \textrm{(\# of subcarriers)} \times \textrm{(\# of antennas)}.
\end{align}
In this paper, we assume that each BS has the same number of dimensions: the $M$-by-$M$ symmetric configuration. The asymmetric case will be discussed in Section~\ref{sec-discussion}. The idea of interference alignment is to design the transmitted vectors so that they are aligned onto a one-dimensional linear subspace at the other BS. Due to the randomness in wireless channels, the transmitted vectors are likely to be linearly independent at the desired BS. Note that for $M=K+1$, the desired signals span a $K$-dimensional linear space while the interference signals only occupy a one-dimensional subspace. Hence, each BS can recover $K$ desired symbols using $K+1$ dimensions.

The performance in the interference-limited regime can be captured by a notion of degrees-of-freedom ($\sf dof$). Here, ${\sf dof \; per \; cell} =\frac{K}{K+1} $, so as $K$ gets large, we can asymptotically achieve interference-free $\sf dof = 1$.
On the other hand, one implementation challenge comes from the overhead of exchanging CSI needed for enabling the IA technique.
The IA scheme requires each user to know its \emph{cross}-channel information to the other BS. While in a time-division-multiplexing system, channels can be estimated using reciprocity, in a frequency-division-multiplexing system, \emph{backhaul cooperation} is required to convey such channel knowledge. Fig.~\ref{fig:uplinkIA} shows a route to obtain the CSI of $\mathbf{G}_{\beta 1}$: $BS\; \beta \rightarrow backhaul \rightarrow BS \; \alpha \rightarrow feedback \rightarrow user \;1 \;of\; cell \; \alpha$. Here $\mathbf{G}_{\beta 1} \in \mathbb{C}^{(K+1) \times (K+1)}$ indicates the cross-channel from user 1 of cell $\alpha$ to BS $\beta$. On the contrary, in the downlink, we show that IA can be applied without backhaul cooperation.

\subsection{Downlink Interference Alignment}
\label{sec:downlinkIA}

\begin{figure}[t]
\begin{center}
{\epsfig{figure=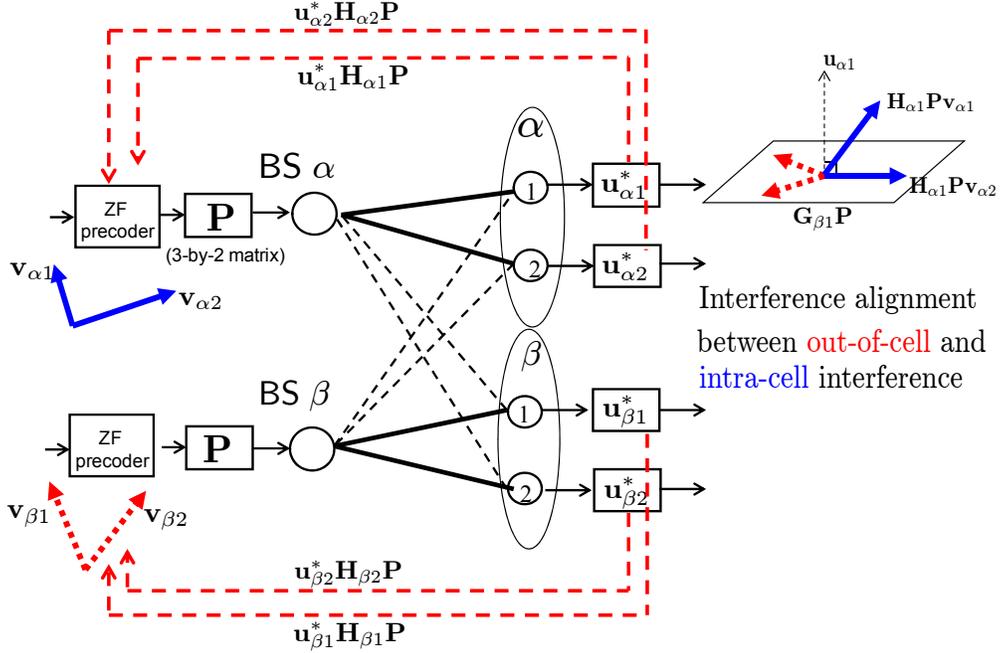, angle=0, width=0.8\textwidth}}
\end{center}
\caption{Downlink interference alignment.
Interference alignment is achieved between out-of-cell and intra-cell interference vectors at multiple users at the same time. Unlike the uplink IA, our downlink IA scheme does not require backhaul cooperation.
} \label{fig:downlinkIA}
\end{figure}

Fig. \ref{fig:downlinkIA} illustrates an example of downlink IA where there are two users in each cell. The uplink-downlink duality says that the $\sf dof$ of the uplink is the same as that of the downlink. Hence, ${\sf dof \; per \; cell} = \frac{K}{K+1}= \frac{2}{3}$. To achieve this, each BS needs to send two symbols (streams) over three dimensions. The idea is similar to that of the uplink IA in a sense that two dimensions are used for transmitting desired signals and the remaining one dimension is reserved for interference signals. However, the method of interference alignment is different.

We first set a 3-by-2 precoder matrix $\mathbf{P}$ at BS $\alpha$ and BS $\beta$, respectively. This spreads two data streams over three-dimensional resources. Next, each user, such as user $1$ in cell $\alpha$, estimates the interference $\mathbf{G}_{\beta 1} \mathbf{P}$ using pilots or a preamble. User $1$ then generates a vector $\mathbf{u}_{\alpha 1}$ that lies in the null-space of the $\mathbf{G}_{\beta 1} \mathbf{P}$. Since the $\mathbf{G}_{\beta 1} \mathbf{P}$ is of dimension 3-by-2, such a vector $\mathbf{u}_{ \alpha 1}$ always exists, and when applied to the received signal, it will null out the out-of-cell interference.

Note that the receive vector $\mathbf{u}_{\alpha 1}$ does not guarantee the cancellation of intra-cell interference from user 2 in the same cell $\alpha$. This is accomplished as follows. In cell $\alpha$, each user feeds back its equivalent channel $\mathbf{u}_{\alpha k}^* \mathbf{H}_{\alpha k}\mathbf{P}$ (obtained after applying the receive vector) to its own BS $\alpha$, where $\mathbf{H}_{\alpha k} \in \mathbb{C}^{3 \times 3}$ indicates the direct-channel from BS $\alpha$ to user $k$ in the cell. BS $\alpha$ then applies an additional zero-forcing precoder formed by the pseudo-inverse of the composite matrix $[\mathbf{u}_{\alpha 1}^* \mathbf{H}_{\alpha 1}\mathbf{P}; \mathbf{u}_{\alpha 2}^* \mathbf{H}_{\alpha 2}\mathbf{P}]$. This zero-forcing precoder guarantees user 2's transmitted signal $\mathbf{H}_{\alpha 1} \mathbf{P} \mathbf{v}_{\alpha 2}$ to lie in the interference space $\mathbf{G}_{\beta 1} \mathbf{P}$. Note that $\mathbf{u}_{\alpha 1}^* (\mathbf{H}_{\alpha 1} \mathbf{P} \mathbf{v}_{\alpha 2}) = 0$.


A series of operations enables interference alignment. Let us call this scheme \emph{zero-forcing IA}. To see this, let us observe the interference plane of user 1 in cell $\alpha$.
Note that there are three interference vectors: two \emph{out-of-cell} interference vectors and one \emph{intra-cell} interference vector. These three vectors are aligned onto a two-dimensional linear subspace. Interference alignment is achieved between out-of-cell and intra-cell interference signals to save one dimension. Similarly, user 2 in the cell can save one dimension. Hence, two dimensions can be saved in total by sacrificing only one dimension. If the number of users is $K$, each cell can save $K$ dimensions by sacrificing one dimension. The loss will become negligible with the increase of $K$, as was seen in the uplink IA.

While the downlink $\sf dof$ is the same as that of the uplink, the way interference is aligned is quite different. Note in Fig.~\ref{fig:uplinkIA} that in uplink IA, interference alignment is achieved among out-of-cell interference vectors only. On the other hand, in downlink IA, interference alignment is achieved between out-of-cell and intra-cell interference vectors at multiple users \emph{at the same time}.

\textbf{Feedback Mechanism:} Note two key system aspects of the technique. First, the exchange of cross-channel information between BSs or between users in different cells is not needed. Each BS can fix precoder $\mathbf{P}$, \emph{independent of channel gains}. Each user can then specify the space orthogonal to the out-of-cell interference signal space. This enables the user to design a zero-forcing receive vector \emph{without knowing the actually transmitted vectors}.
Each user then feeds back the equivalent channel  $\mathbf{u}_{\alpha k} \mathbf{H}_{\alpha k}\mathbf{P}$ and the BS forms the zero-forcing transmit vectors \emph{only with the feedback of the equivalent channels}. Hence, the scheme requires only within-a-cell feedback mechanism. This is in stark contrast to the uplink IA which requires backhaul cooperation between different BSs.

Secondly, while feedback is required from the user to the BS, this feedback is the same as the feedback used for standard multi-user MIMO techniques. The only difference is that in downlink IA, two cascaded precoders are used and the receive vector of each user is chosen as a null vector of out-of-cell interference signal space. As a result, the scheme can be implemented with little change to an existing cellular system supporting multi-user MIMO.

\subsection{Performance and Limitations}

\begin{figure}[t]
\begin{center}
{\epsfig{figure=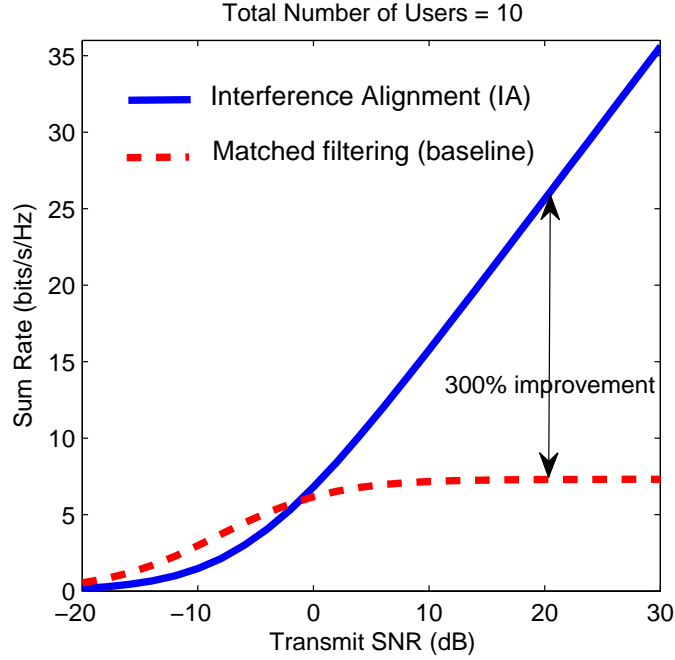, angle=0, width=0.6\textwidth}}
\end{center}
\caption{Performance of downlink interference alignment for a two-isolated cell layout with a 4-by-4 antenna configuration ($M=4$). An opportunistic scheduler is employed to choose a set of 3 users out of 10 such that the sum rate is maximized.
} \label{fig:Performance2cellonly}
\end{figure}

Fig.~\ref{fig:Performance2cellonly} shows the sum-rate performance for downlink zero-forcing IA in a two-isolated cell layout where $M=4$ (e.g., a 4-by-4 antenna configuration). As a baseline scheme, we use a \emph{matched filter receiver}: one of the standard multi-user MIMO techniques \cite{Pan:IterativeMF,Gesbert:MUMIMO}. The scheme uses the dominant left-singular vector of the direct-channel as a receive vector. We assume a zero-forcing vector at the transmitter to null out intra-cell interference. Nulling intra-cell interference is important as its power has the same order as the desired signal power.
\begin{align}
\label{eq-MFreceiveweight}
& \mathbf{u}_{\alpha k}^{\sf MF} = \textrm{a maximum left-singular vector of } \mathbf{H}_{\alpha k}, \\
\label{eq-ZFtransmitweight}
& \mathbf{v}_{\alpha k}^{\sf ZF}= k\textrm{th normalized column of } \mathbf{H} (\mathbf{H}\mathbf{H}^*)^{-1}, \; \mathbf{H}:= \left[
  \begin{array}{c}
  \vdots \\
   \mathbf{u}_{\alpha k}^{{\sf MF}*}  \mathbf{H}_{\alpha k}  \\
     \vdots \\
  \end{array}
\right].
\end{align}
Note that the matched filter receiver maximizes beam-forming gain but it ignores the interference signal space. Also notice that the receive-and-transmit vectors are interconnected, i.e., a receiver vector can be updated as a function of a transmit vector and vice versa. One way to compute the transmit-and-receive vectors is to employ an iterative algorithm \cite{Pan:IterativeMF,Gesbert:MUMIMO}. We call this scheme \emph{iterative matched filtering}.
See Appendix~\ref{appen:IterativeMF} for further details. In Fig.~\ref{fig:Performance2cellonly}, we assume no iteration for fair comparison of CSI overhead.
\begin{figure}[t]
\begin{center}
{\epsfig{figure=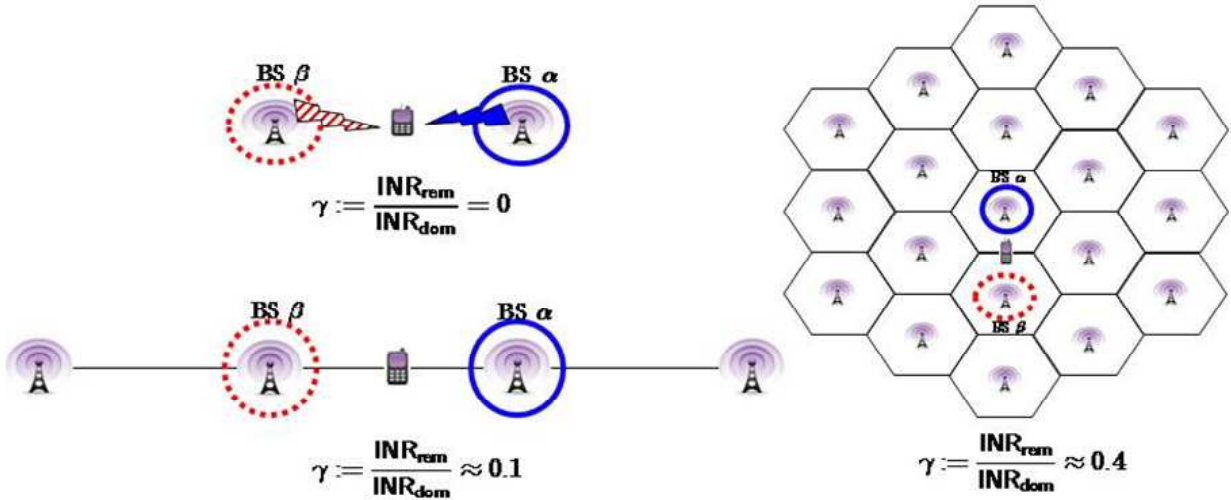, angle=0, width=1.0\textwidth}}
\end{center}
\caption{Different layouts in a downlink cellular system. A parameter $\gamma$ indicates the relative strength of the interference power from a dominant interferer to the remaining interference power (summed from the other BSs).} \label{fig:gamma}
\end{figure}
An opportunistic scheduler~\cite{Tse:Book} is employed to choose a set of 3 users out of 10 such that the sum rate is maximized.
We also consider uncoordinated schedulers, i.e., scheduling information is not exchanged between different BSs.

One can clearly see that the zero-forcing IA provides significant (asymptotically optimum for large $\sf SNR$) performance gain for the two-isolated-cell case, as there are no residual interferers. However, for realistic multi-cellular environments, the performance may not be very good due to the remaining interferers. In order to take multi-cellular environments into account, we introduce a parameter $\gamma$ that captures the relative strength of the interference power from a dominant interferer to the remaining interference power (summed from the other BSs):
\begin{align}
\label{eq-gamma}
\gamma:= \frac{\sf INR_{rem}}{\sf INR_{dom}},
\end{align}
where $\sf INR_{dom}$ and $\sf INR_{rem}$ denote the ratios of the dominant and aggregate interference power over the noise power, respectively. Note that by adapting $\gamma$, one can cover arbitrary mobile location and cellular layouts.

While, at one extreme ($\gamma =0$), the zero-forcing IA provides significant performance, at the other extreme ($\gamma \gg 1$), the scheme may not be good as it completely loses receive beam-forming gain (the zero-forcing IA receiver is independent of the direct-channel since it depends only on the interference space). In this case, one can expect that matched filtering will perform much better than the IA scheme. This motivates the need for developing a new IA technique that can balance the degrees-of-freedom gain with the matched-filtered power gain depending on the value of $\gamma$.

\section{Proposed New IA Scheme}
\label{sec-proposedscheme}

The zero-forcing IA and matched filtering schemes remind us of a conventional zero-forcing receiver and a matched filter receiver in a point-to-point channel with colored noise. So it is natural to think of a unified technique like the standard MMSE receiver. However, in our cellular context, a straightforward design of an MMSE receiver usually requires the knowledge of transmitted vectors from the other cell. Moreover, a chicken-and-egg problem arises between different cells, due to the interconnection of the transmit-and-receive vector pairs.


In order to \emph{decouple} the vector design between cells, we consider uncoordinated systems, i.e., transmit vector information is not exchanged between different cells. Under this assumption, a goal is to mimic an MMSE receiver. The idea is to \emph{color} an interference signal space by using two cascaded precoders, one of which is a \emph{fixed} precoder $\mathbf{\bar{P}}$ located at the front-end. With the fixed precoder, we can color the interference space, to some extent, to be independent of actually transmitted vectors. To see this, consider the covariance matrix of interference-plus-noise:
\begin{align}
\mathbf{\Phi}_{ k} = (1 + {\sf INR_{rem} }) \mathbf{I} +
\frac{ \sf SNR }{S}  (\mathbf{G}_{\beta k} \mathbf{\bar{P}} \mathbf{B}_{\beta } \mathbf{B}_{\beta }^* \mathbf{\bar{P}}^* \mathbf{G}_{\beta k}^* ),
\end{align}
where $S$ is the total number of streams assigned to the scheduled users in the cell ($S \leq M$) and $\mathbf{B}_{\beta}$ indicates the zero-forcing precoder of a dominant interferer (BS $\beta$): $\mathbf{B}_{\beta} = [\mathbf{v}_{\beta 1}, \cdots, \mathbf{v}_{\beta S}] \in \mathbb{C}^{M \times S}$. Assume that the aggregate interference except the dominant interference is white Gaussian\footnote{To be more accurate, we may consider two or three dominant interferers for an actual realization. See Section \ref{sec-discussion} for details.}. Without loss of generality, we assume that Gaussian noise power is normalized to 1. Assume the total transmission power is equally allocated to each stream.

We control the coloredness of interference signals by differently weighting the last $(M-S)$ columns of $\mathbf{\bar{P}}$ with a parameter $\kappa$ ($0 \leq \kappa  \leq 1$):
\begin{align}
 \mathbf{\bar{P}} = \left[ \mathbf{f}_1, \cdots, \mathbf{f}_{S}, \kappa \mathbf{f}_{S+1}, \cdots,  \; \kappa \mathbf{f}_M \right] \in \mathbb{C}^{M \times M},
\end{align}
where $\left[ \mathbf{f}_1, \cdots, \mathbf{f}_{M} \right]$ is a unitary matrix.
Since we consider uncoordinated systems, $\mathbf{B}_{\beta}$ is unknown. Hence, we use the expectation of the covariance matrix over $\mathbf{B}_{\beta}$:
\begin{align}
\mathbf{\bar{\Phi}}_{ k}:= \mathbb{E} [\mathbf{\Phi}_{ k}] =
 (1 + {\sf INR_{rem} }) \mathbf{I} +
\frac{ \sf SNR }{S}  (\mathbf{G}_{\alpha k} \mathbf{\bar{P}} \mathbf{\bar{P}}^* \mathbf{G}_{\alpha k}^* ),
\end{align}
where we assume that each entry of $\mathbf{B}_{\beta}$ is i.i.d. $\mathcal{CN}(0, \frac{1}{S} )$.

Two extreme cases give insights into designing $\kappa$.
When the residual interference is negligible, i.e., $\gamma \ll 1$, the scheme should mimic the zero-forcing IA, so  $\mathbf{\bar{P}}$ should be rank-deficient, i.e., $\kappa =0$. Note in this case that the null space of the interference signals can be specified, independent of $\mathbf{B}_{\beta}$. As a result, the expected covariance matrix acts as the actual covariance matrix to induce the solution of the zero-forcing IA. At the other extreme ($\gamma \gg 1$), the scheme should mimic matched filtering. This motivates us to choose a unitary matrix $\mathbf{\bar{P}}$. One way for smoothly sweeping between the two cases is to set:
\begin{align}
\label{eq-KAPPA}
\kappa = \min \left( \sqrt{ \gamma } , 1 \right).
\end{align}
Note that for $\gamma \ll 1$, $\kappa \approx 0$ and for $\gamma \gg 1$, $\kappa$ is saturated as 1.

Considering system aspects, however, the $\kappa$ needs to be carefully chosen. In the above choice, the $\kappa$ varies with mobile location, since $\sf INR_{rem}$ is a function of mobile location. This can be undesirable because it requires frequent adaptation of BS precoder which supports users from the cell center to the cell edge.
Therefore, we propose to fix $\kappa$. For example, we can fix $\kappa$ based on the case of $\sf SNR=20$ dB, a cell-edge mobile location, and a fixed network layout, e.g.,
$\kappa \approx 0.34$ for the linear cell layout and $\kappa \approx 0.64$ for the 19 hexagonal wrap-around cell layout (See Fig.~\ref{fig:gamma}).

With $\mathbf{\bar{\Phi}}_k$, we then use the standard formula of an MMSE receiver.
Similar to the iterative matched filtering technique, we also employ an iterative approach to compute transmit-and-receive vector pairs.

\textbf{$<$Proposed New IA Scheme$>$}
\begin{enumerate}
  \item (\emph{Intialization}): Each user initializes its receive vector as follows: $\forall k \in \{1,\cdots, K \},$
  \begin{align}
  \begin{split}
   \mathbf{u}_{\alpha k}^{(0)} = \textrm{normalization} \left\{
 \mathbf{\bar{\Phi}}_{ k}^{-1}  \mathbf{H}_{\alpha k} \mathbf{\bar{P}} \mathbf{v}_{\alpha k}^{(0)} \right \},
  \end{split}
  \end{align}
  where we set $\mathbf{v}_{\alpha k}^{(0)}$ as a maximum eigenvector of $\mathbf{\bar{P}}^*  \mathbf{H}_{\alpha k}^*
 \mathbf{\bar{\Phi}}_{ k}^{-1}
\mathbf{H}_{\alpha k} \mathbf{\bar{P}}$ to initially maximize beam-forming gain.
  Each user then feeds back the equivalent channel $\mathbf{u}_{\alpha k}^{(0)*} \mathbf{H}_{\alpha k} \mathbf{\bar{P}}$ to its own BS. With this feedback information, the BS computes zero-forcing transmit vectors: $\forall k$
  \begin{align*}
      \begin{split}
      \mathbf{v}_{\alpha k}^{(1)} =  k\textrm{th normalized column of } \mathbf{H}^{(1)*} (\mathbf{H}^{(1)} \mathbf{H}^{(1)*})^{-1},
      \end{split}
  \end{align*}
  where
  \begin{align}
  \begin{split}
  \mathbf{H}^{(1)}:= \left[
  \begin{array}{c}
  \vdots \\
   \mathbf{u}_{\alpha k}^{(0)*}  \mathbf{H}_{\alpha k} \mathbf{\bar{P}}  \\
     \vdots \\
  \end{array}
\right]
   \end{split}
   \end{align}.
  \item (\emph{Opportunistic Scheduling}): The BS finds $A^*$ such that
  \begin{align*}
A^{*}  =   \arg \max_{ A \in \mathcal{K} } \sum_{k \in A } \log \left( 1 + \frac{ \frac{ {\sf SNR} }{S} || \mathbf{u}_{\alpha k}^{(0)*} \mathbf{H}_{\alpha k}  \mathbf{v}_{\alpha k}^{(1)}    ||^2  }{1 +  {\sf INR_{rem} }
      } \right),
\end{align*}
where $\mathcal{K}$ is a collection of subsets $ \subset \{1,\cdots,K\}$ that has cardinality $|\mathcal{K}| = \binom{K}{S}$.
  \item (\emph{Iteration}): For $A^*$, we iterate the following. The BS informs each user of $\mathbf{v}_{\alpha k}^{(i)}$ via precoded pilots. Each user updates the receive vector as follows:
 \begin{align*}
        \mathbf{u}_{\alpha k}^{(i)} = \textrm{normalization} \left\{
 \mathbf{\bar{\Phi}}_{ k}^{-1}  \mathbf{H}_{\alpha k} \mathbf{\bar{P}} \mathbf{v}_{\alpha k}^{(i)} \right \}.
 \end{align*}
 Each user then feeds back the updated equivalent channel to its own BS. With this feedback information, the BS computes zero-forcing transmit vectors $\mathbf{v}_{\alpha k}^{(i+1)}$.
\end{enumerate}

\textbf{Remarks:}
Although users can see out-of-cell interference, the scheduler at BS cannot compute it. Hence, we assume that the scheduler makes a decision assuming no dominant interference. Note that the denominator inside the logarithmic term contains only noise and residual interference. To reduce CSI overhead, we assume that a scheduler decision is made before the \emph{iteration} step.

In practice, we may not prefer to iterate, since it requires more feedback information. Note that the feedback overhead is exactly the same as that of iterative matched-filtering (baseline). The only difference is that we use the fixed precoder $\mathbf{\bar{P}}$ and the MMSE-like receiver employing the $\mathbf{\bar{\Phi}}_{ k}$. \emph{This requires very little change to an existing cellular system supporting multi-user MIMO}.

\section{Simulation Results}
\label{sec-simulation}

Through simulations, we evaluate the performance of the proposed scheme for downlink cellular systems. We consider one of the possible antenna configurations in the 4G standards~\cite{80216m,3GPPLTE}: $4$ transmit and 4 receive antennas. To minimize the change to the existing 4G systems, we consider using only antennas for the multiple dimensions, i.e., $M=4$. We focus on three different cellular layouts, illustrated in Fig.~\ref{fig:gamma}. We consider a specific mobile location (the mid-point between two adjacent cells), as the cell-edge throughput performance is of our main interest. We use the standard ITU-Ped path-loss model, with i.i.d. Rayleigh fading components for each of the antenna.

\begin{figure}[t]
\centering
\subfigure[]{
\includegraphics[width=2.6in]{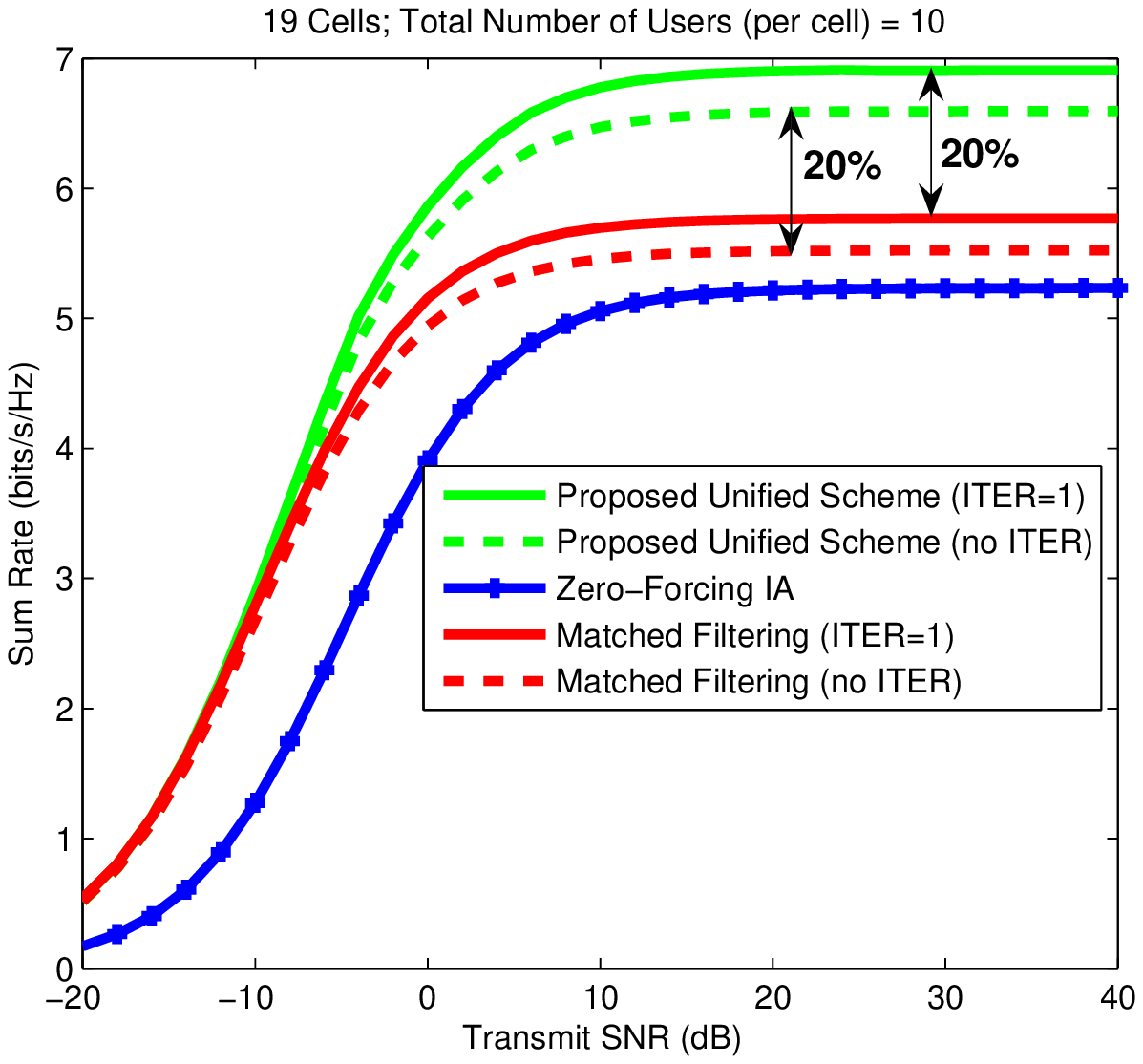}}
\hspace{1in}
\subfigure[]{
\includegraphics[width=2.6in]{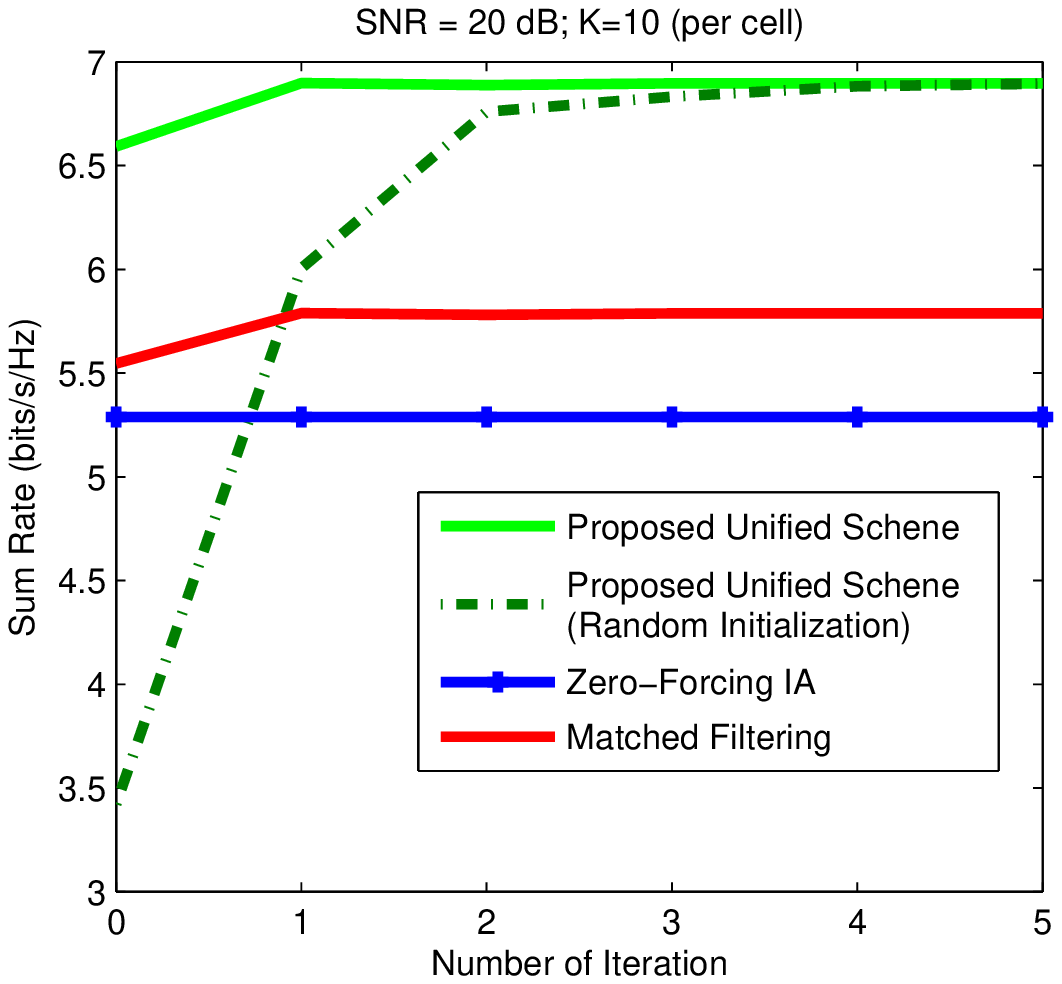}}
\caption{The sum-rate performance for a 19 hexagonal cell layout where the number $K$ of users per cell is 10 and the number $S$ of streams is 3: $(a)$ as a function of transmit $\sf SNR$; $(b)$ as a function of the number of iteration.
} \label{fig:19hexa}
\end{figure}

Fig.~\ref{fig:19hexa} shows the throughput performance for 19 hexagonal cellular systems where $\gamma \approx 0.4$. We consider $K=10$ and $S=3$. We find through simulations that using three streams provides the best performance for a practical number of users per cell (around 10). See Appendix~\ref{appendix:NumberofStreams} for further details. Note that the zero-forcing IA scheme is worse than the matched filtering (baseline). This implies that when $\gamma \approx 0.4$ (residual interference is not negligible), boosting power gain gives better performance than mitigating dominant out-of-cell interference. However, the proposed unified IA technique outperforms both of them for all regimes. It gives approximately 20\% throughput gain when ${\sf SNR} = 20$ dB.

We also investigate the convergence of the proposed scheme. Note in Fig.~\ref{fig:19hexa}$(b)$ that the proposed scheme converges to the limits very fast, i.e., even one iteration is enough to derive most of the asymptotic performance gain. This means that additional iterations provide marginal gain, while requiring a larger overhead of CSI feedback. Another observation is that the converged limits of the proposed technique is invariant to the initial values of transmit-and-receive vectors. Note that random initialization induces the same limits as that of our carefully chosen initial values, but it requires more iterations to achieve the limits. Therefore, initial values need to be carefully chosen to minimize the overhead of CSI feedback.



\begin{figure}[t]
\begin{center}
{\epsfig{figure=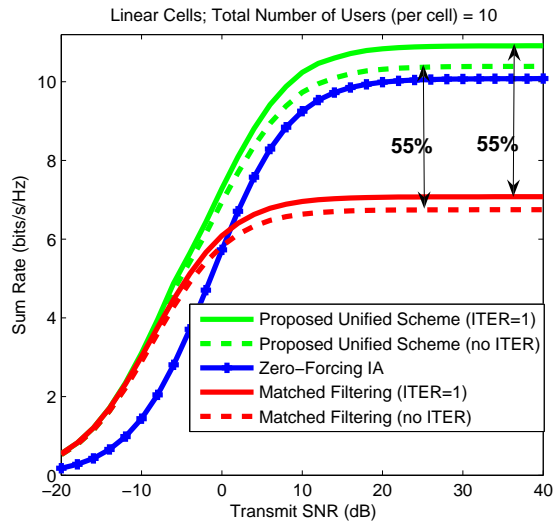, angle=0, width=0.5\textwidth}}
\end{center}
\caption{The sum-rate performance for a linear cell layout as a function of $\sf SNR$. We consider $K=10$ (per cell) and $S=3$.} \label{fig:linearcell}
\end{figure}

Fig.~\ref{fig:linearcell} shows throughput performance for a linear cell layout. In this case, the residual interference is significantly reduced at $\gamma \approx 0.1$, so mitigating dominant out-of-cell interference improves the performance more significantly than beam-forming does. The gain of the proposed scheme over the matched filtering is significant, i.e., approximately 55\% in the high $\sf SNR$ regime of interest. Notice that a crossover point between the zero-forcing IA and the matched filtering occurs at around ${\sf SNR}=0$ dB. The benefit of the zero-forcing IA is substantial.




\section{Macro-Pico Cellular Networks}
\label{sec-Application}

\begin{figure}[t]
\begin{center}
{\epsfig{figure=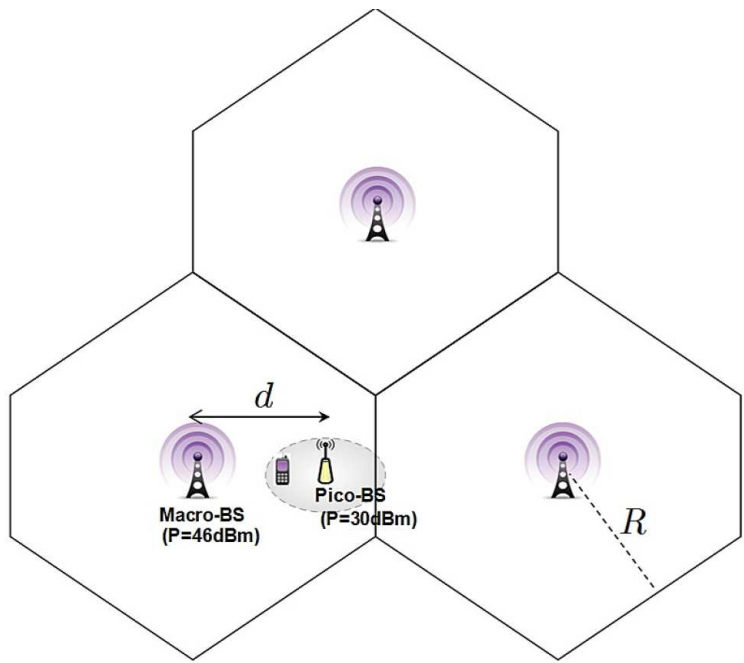, angle=0, width=0.6\textwidth}}
\end{center}
\caption{Macro-pico cellular networks. The pico-user can see significant interference from the nearby macro-BS. The interference problem can be further aggravated when the pico-BS is close to the nearby macro-BS (small $d$) and the power levels of the two BSs are quite different.} \label{fig:macro-pico}
\end{figure}

We have observed that our scheme shows promise especially when
dominant interference is much stronger than the remaining interference, i.e., $\gamma \ll 1$. Such scenario occurs often in heterogeneous networks~\cite{Heterogeneous} which use a mix of macro, pico, femto, and relay BSs to enable flexible and low-cost deployment.
In this section, we focus on a scenario of the macro-pico cell deployment, illustrated in Fig.~\ref{fig:macro-pico}.

As shown in the figure, suppose that pico-BS is deployed at a distance $d$ from the nearby macro-BS and a user is connected to the pico-BS.
The pico-user can then see significant interference from the nearby macro-BS, and this interference can be much stronger than the aggregated interference from the remaining macro-BSs, especially when $d$ is small. The interference problem can be further aggravated due to
range extension techniques\footnote{Range extension extends the footprint of pico-cells by allowing more users to connect even if users do not see the pico-BS as the strongest downlink received power. The purpose for this is to better utilize cell-splitting and maximize cell offloading gain.}~\cite{Heterogeneous} and the disparity between the transmit power levels of the macro-BS and the pico-BS. This motivates the need for intelligent interference management techniques. We show that our IA scheme can resolve this problem to provide substantial gain.

To show this, we evaluate the throughput performance of pico-users in the simple scenario shown in Fig.~\ref{fig:macro-pico}. We assume the 19 hexagonal wrap-around cellular layout, and on top of it we deploy one pico-BS. Based on~\cite{Heterogeneous}, we consider the power levels of $46$ dBm and $30$ dBm for the macro-BS and the pico-BS, respectively, so the difference is $16$ dB. Consistent with previous simulation setups, we consider a specific mobile location where the downlink received power from the pico-BS is the same as that from the nearby macro-BS. Due to the disparity of the power levels, the pico-users are closer to the pico-BS. We assume a $4$-by-$4$ antenna configuration where $M=4$.


\begin{figure}[t]
\centering
\subfigure[]{
\includegraphics[width=2.6in]{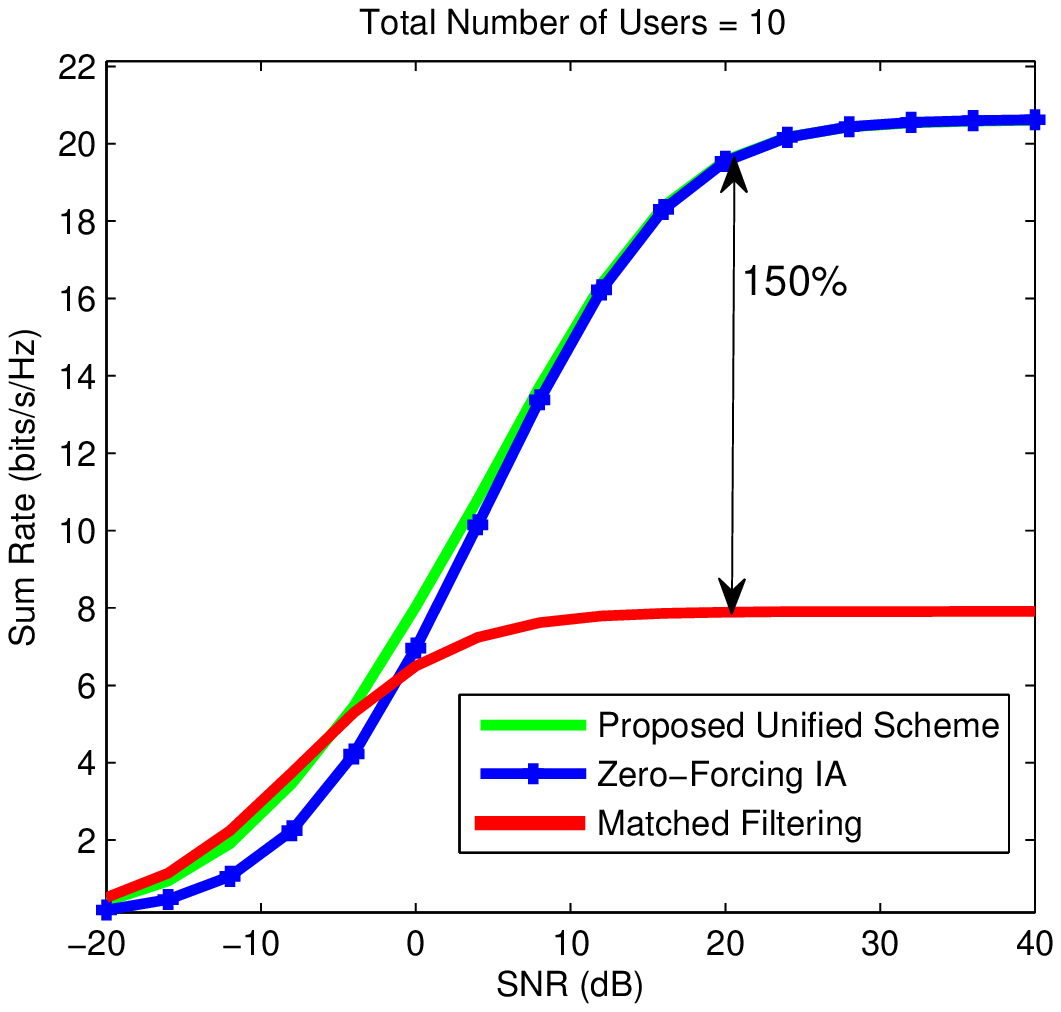}}
\hspace{1in}
\subfigure[]{
\includegraphics[width=2.6in]{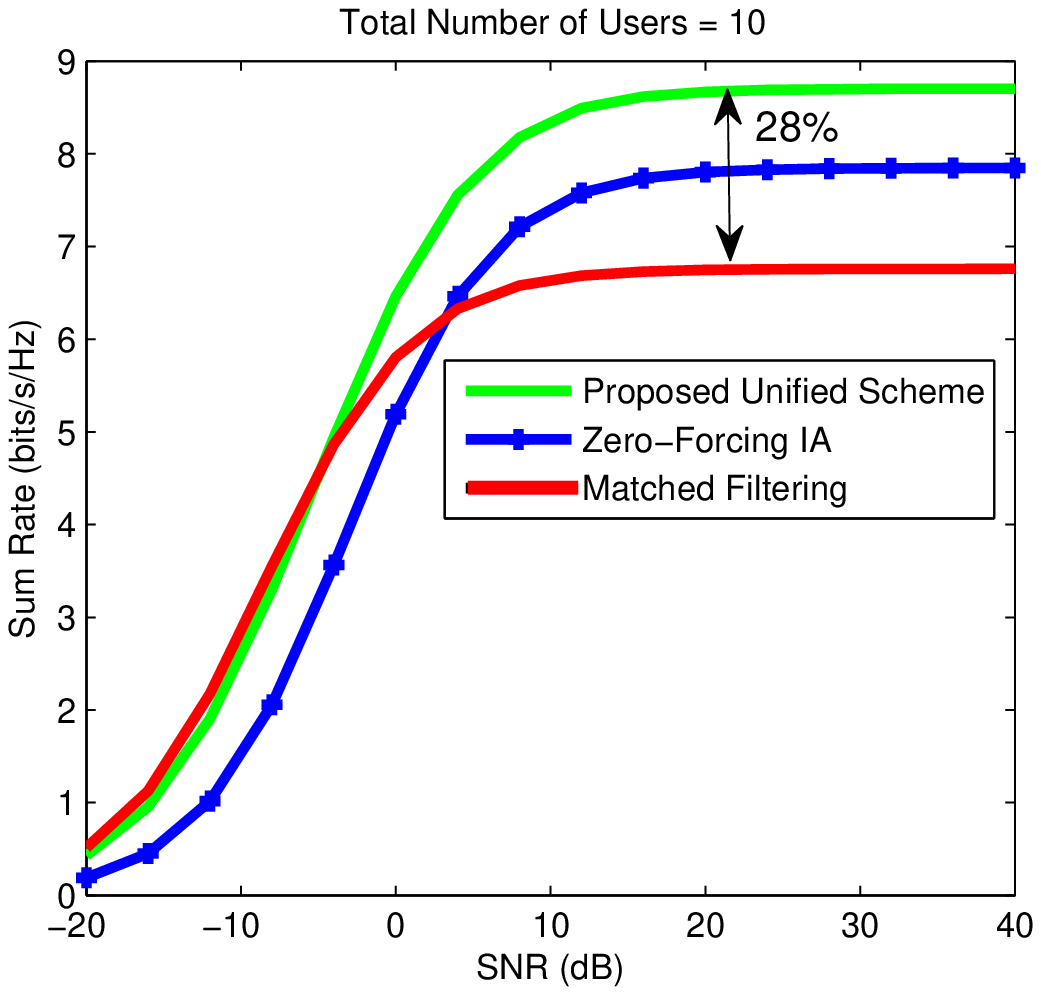}}
\caption{The sum-rate performance for macro-pico cell layout (on top of 19 wrap-around macro cells): $(a)$ $\frac{d}{R}=0.5$; $(b)$ $\frac{d}{R}=1$. The number $K$ of users per cell is 10; the number $S$ of streams is 3; and no iteration is performed.
} \label{fig:macro-pico_SNR}
\end{figure}

Fig.~\ref{fig:macro-pico_SNR} shows the throughput performance of the pico-users as a function of $\sf SNR$. We assume that $K=10$, $S=3$ and no iterations. We employ the opportunistic scheduler to choose the best 3 users out of 10. Fig.~\ref{fig:macro-pico_SNR} $(a)$ considers the case of $\frac{d}{R}=0.5$ where pico-users are significantly interfered with by the nearby macro-BS. In this case, as one can expect, our IA scheme provides significant gain of $150$\% over the matched filtering, similar to the two-isolated cell case. In Fig.~\ref{fig:macro-pico_SNR} $(b)$, we also consider the case of $\frac{d}{R}=1$ where the minimum gain of our scheme is expected. Even in this worst case, our proposed scheme gives approximately $28$\% gain over the matched filtering.

Recall that in this simulation we consider the specific mobile location where the downlink received power from the two BSs are the same. In fact, this is a conservative case. As mentioned earlier, the use of the range extension technique expands the footprint of pico-cells and therefore aggravates the interference problem. One can expect a larger gain of our IA scheme when range extension is employed.

\begin{figure}[t]
\begin{center}
{\epsfig{figure=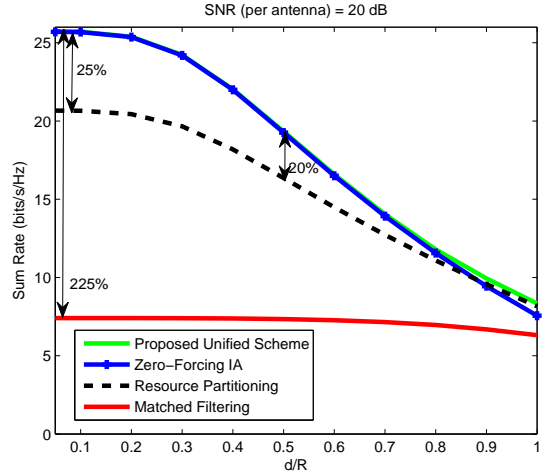, angle=0, width=0.5\textwidth}}
\end{center}
\caption{Comparison to resource partitioning. The sum-rate performance as a function of $\frac{d}{R}$ for ${\sf SNR} = 20$ dB.} \label{fig:macro-pico_comparison}
\end{figure}

\textbf{Comparison to Resource Partitioning:} In this scenario, as an alternative to our IA scheme, one may consider resource partitioning to resolve the interference problem. This is because unlike the conventional macro cellular networks containing many neighboring cells, this macro-pico network scenario has a fewer number of dominant interferers, thus making resource coordination simpler.
For example, we can use a frequency reuse of $\frac{1}{2}$ for the scenario in Fig.~\ref{fig:macro-pico}. However, resource partitioning requires \emph{explicit} coordination of frequency resources which can increase the control channel overhead. On the contrary, our IA scheme does not require explicit coordination, as it adapts only the number of streams under frequency reuse of $1$.
In addition to this implementation advantage, our scheme shows respectable gain over resource partitioning. Fig.~\ref{fig:macro-pico_comparison} shows the throughput performance as a function of $\frac{d}{R}$ when ${\sf SNR}=20$ dB and $K=10$. We use $S=3$ for the IA schemes and the matched filtering, while for resource partitioning we optimize the number of streams to plot the best performance curve. Notice that our scheme gives approximately $20$\% gain for $\frac{d}{R}=0.5$. The smaller ratio of $\frac{d}{R}$, the larger the gain, while for large $\frac{d}{R}$, the gain becomes marginal.

\section{Extension}
\label{sec-discussion}

\subsection{Asymmetric Antenna Configuration}
As one natural extension, we consider asymmetric antenna configuration where the BSs are equipped with more antennas. A slight modification of our technique can cover this case.
Consider $M$-by-$N$ antenna configuration where $M>N$. Compared to the symmetric case, the only difference is that the number of streams is limited by the number $N$ of receive antennas, i.e., $S \leq N$. Other operations remain the same.

Specific operations are as follows. Each BS sets the precoder $\mathbf{\bar{P}}$ as follows:
\begin{align}
\mathbf{\bar{P}} = [\mathbf{f}_1, \cdots, \mathbf{f}_S, \kappa \mathbf{f}_{S+1}, \cdots, \kappa \mathbf{f}_M] \in \mathbb{C}^{M \times M},
\end{align}
where $0 \leq \kappa \leq 1$. Notice that $S \leq N$. Each user computes the expected covariance matrix by averaging over the transmitted signals from the other cell and then applies the standard MMSE formula for a receive vector. The BS then computes the zero-forcing transmit vectors with the feedback information. These steps can then be iterated.

While our technique can be extended to any antenna configuration, interpretation needs to be carefully made for some cases. For example, consider 4-by-2 antenna configuration in a two-cell layout. Our scheme allows each BS to send one stream out of two and therefore each user sees only one interference vector from the other cell. This induces no interference alignment.
Even in this configuration, however, interference alignment can be achieved if multiple subcarriers are incorporated. This will be discussed in the following section.

\subsection{Using Subcarriers}
Recall in our simulations that only antennas are employed to generate multiple dimensions. We can also increase $M$ by using multiple subcarriers, thereby improving performance as the dimension reserved for interference signals becomes negligible with the increase of $M$. For example, we can create 8-by-4 configuration by using two subcarriers in a 4-by-2 antenna configuration.

Interestingly, unlike the 4-by-2 configuration, this 8-by-4 configuration enables interference alignment.
To see this, consider a two-cell layout where each cell has three users. Our scheme allows each BS to transmit three streams out of four and thus each user sees five interfering vectors in total: three \emph{out-of-cell} and two \emph{intra-cell} interfering vectors. Notice the five interfering vectors are aligned onto a three dimensional linear subspace, thereby achieving interference alignment.





\subsection{Open-Loop Multi-User MIMO}
Since the feedback mechanism of our scheme is the same as that of standard multi-user MIMO techniques, any CSI feedback reduction scheme used for standard techniques can also be applied to our proposed scheme.
For example, an open-loop multi-user MIMO technique can be easily applied to our scheme. Our scheme has only two differences: (1) each BS employs two cascaded precoders, including a fixed precoder $\mathbf{\bar{P}}$; (2) each user employs an MMSE-like receiver using $\mathbf{\bar{\Phi}}_k$.


\subsection{Multiple Interferers}

Our IA technique removes the interference from a single dominant interferer. A slight modification can be made to cope with multiple dominant interferers. For example, consider a 19 hexagonal cell layout in Fig.~\ref{fig:gamma} and suppose that mobiles are located at the middle point of three neighboring BSs. In this case, mobiles see two dominant interferers. One simple way is to take multiple dominant interferers into account in the process of computing the expected covariance matrix. Specifically, we can use:
\begin{align}
\begin{split}
\mathbf{\bar{\Phi}}_{k}&:=
\mathbb{E} \left[ (1 + {\sf INR_{rem}}) \mathbf{I} + \frac{\sf SNR}{S}
\mathbf{G}_{ \beta k} \mathbf{\bar{P}} \mathbf{B}_{\beta} \mathbf{B}_{\beta}^* \mathbf{\bar{P}}^* \mathbf{G}_{\beta k}^{*}  + \frac{\sf SNR}{S}
\mathbf{G}_{ \gamma k} \mathbf{\bar{P}} \mathbf{B}_{\gamma} \mathbf{B}_{\gamma}^* \mathbf{\bar{P}}^* \mathbf{G}_{\gamma k}^{*}  \right] \\
& = (1 + {\sf INR_{rem}}) \mathbf{I} + \frac{\sf SNR}{S}
\mathbf{G}_{ \beta k} \mathbf{\bar{P}} \mathbf{\bar{P}}^* \mathbf{G}_{\beta k}^{*}  + \frac{\sf SNR}{S}
\mathbf{G}_{ \gamma k} \mathbf{\bar{P}} \mathbf{\bar{P}}^* \mathbf{G}_{\gamma k}^{*},
\end{split}
\end{align}
where $\mathbf{G}_{\beta k}$ denotes cross-link channel from BS $\beta$ to user $k$ in cell $\alpha$ and $\mathbf{B}_{\beta}$ indicates the zero-forcing precoder of BS $\beta$, and we use similar notation ($\mathbf{G}_{\gamma k}, \mathbf{B}_{\gamma}$) for cell $\gamma$.  We further assume that each entry of $\mathbf{B}_{\beta}$ and $\mathbf{B}_{\gamma}$ is i.i.d. $\mathcal{CN}(0,\frac{1}{S})$.

\subsection{Optimization of $\kappa$}
Our proposed scheme employs a parameter $\kappa$ in constructing the precoder $\mathbf{\bar{P}}$. We have considered one particular choice of (\ref{eq-KAPPA}), and simulation results are based on this choice. However, the performance can be improved by optimizing $\kappa$. It could be future work to find the optimum $\kappa$ for different cellular layouts.

\section{Conclusion}
We have observed that the zero-forcing IA scheme is analogous to the zero-forcing receiver, and the iterative matched-filtering technique corresponds to the conventional matched-filter receiver. Based on this observation, we proposed a unified IA technique similar to an MMSE receiver that outperforms both techniques for all values of $\gamma$, where the power of the dominant interferer may be much greater or smaller than the power of the remaining aggregate interference.

Of practical importance is the fact that our proposed scheme can be implemented with small changes to an existing cellular system supporting multi-user MIMO, as it requires only a localized \emph{within-a-cell} feedback mechanism.
This technique can be extended to asymmetric antenna configurations, scenarios with more than one dominant interferer, and low CSI schemes such as open-loop MU-MIMO. Our technique also shows even greater performance gains for macro-pico cellular networks where the dominant interference is much stronger than the remaining interference.


\appendices

\section{Iterative Matched Filtering (Baseline)}
\label{appen:IterativeMF}

%

We compute the transmit-and-receive vector pairs using an iterative algorithm \cite{Pan:IterativeMF,Gesbert:MUMIMO}. We describe the algorithm combined with opportunistic scheduler.

\begin{enumerate}
  \item (\emph{Intialization}): Each user initializes a receive vector so as to maximize beam-forming gain: $\forall k \in \{ 1,\cdots, K\}$,
  \begin{align}
  \begin{split}
 \mathbf{u}_{\alpha k}^{(0)} = \textrm{a maximum left-singular vector of } \mathbf{H}_{\alpha k}.
  \end{split}
  \end{align}
   Each user then feeds back the equivalent channel $\mathbf{u}_{\alpha k}^{(0)*} \mathbf{H}_{\alpha k}$ to its own BS.
   With this feedback information, the BS computes zero-forcing transmit vectors:
         $\forall k$,
   \begin{align}
      \begin{split}
      \mathbf{v}_{\alpha k}^{(1)} =  k\textrm{th normalized column of } \mathbf{H}^{(1)*} (\mathbf{H}^{(1)} \mathbf{H}^{(1)*})^{-1},
      \end{split}
   \end{align}
   where
   \begin{align}
    \begin{split}
       \mathbf{H}^{(1)}:= \left[
  \begin{array}{c}
  \vdots \\
   \mathbf{u}_{\alpha k}^{(0)*}  \mathbf{H}_{\alpha k}  \\
     \vdots \\
  \end{array}
\right]
      \end{split}
      \end{align}
  \item (\emph{Opportunistic Scheduling}): The BS finds $A^*$ such that
\begin{align}
A^{*}  =   \arg \max_{ A \in \mathcal{K} } \sum_{k \in A } \log \left( 1 + \frac{ \frac{ {\sf SNR} }{S} || \mathbf{u}_{\alpha k}^{(0)*} \mathbf{H}_{\alpha k}  \mathbf{v}_{\alpha k}^{(1)} ||^2  }{1 +  {\sf INR_{dom} }   + {\sf INR_{rem} }
      } \right).
\end{align}
where $\mathcal{K}$ is a collection of subsets $ \subset \{1,\cdots,K\}$ that has cardinality $|\mathcal{K}| = \binom{K}{S}$.
  \item (\emph{Iteration}): For $A^*$, we iterate the following. The BS informs each user of $\mathbf{v}_{\alpha k}^{(i)}$ via precoded pilots.
  Each user updates the receive vector:
      \begin{align}
  \begin{split}
  \mathbf{u}_{\alpha k}^{(i)} = \textrm{normalization} \left\{
 \mathbf{H}_{\alpha k} \mathbf{v}_{\alpha k}^{(i)} \right\}, k \in A^*.
  \end{split}
  \end{align}
 Each user then feeds back the updated equivalent channel $\mathbf{u}_{\alpha k}^{(i)*} \mathbf{H}_{\alpha k}$ to its own BS.
  With this feedback information, the BS computes zero-forcing transmit vectors:
      \begin{align}
      \begin{split}
      \mathbf{v}_{\alpha k}^{(i+1)} =  k\textrm{th normalized column of } \mathbf{H}^{(i+1)*} (\mathbf{H}^{(i+1)} \mathbf{H}^{(i)*})^{-1},
      \end{split}
      \end{align}
      where
      \begin{align}
      \begin{split}
      \mathbf{H}^{(i+1)}:= \left[
  \begin{array}{c}
  \vdots \\
   \mathbf{u}_{\alpha k}^{(i)*}  \mathbf{H}_{\alpha k}  \\
     \vdots \\
  \end{array}
\right].
      \end{split}
      \end{align}
\end{enumerate}

\textbf{Remarks:} Although users can see out-of-cell interference, the scheduler at BS cannot compute it. We assume the scheduler uses the average power of the dominant interference. Note that the denominator inside the logarithmic term contains noise, dominant interference and residual interference. To reduce CSI overhead, we assume a scheduler decision is made before an iteration. In practice, we may prefer not to iterate, since it requires more feedback information.

\section{Discussion on the Number of Streams}
\label{appendix:NumberofStreams}
\begin{figure}[t]
\begin{center}
{\epsfig{figure=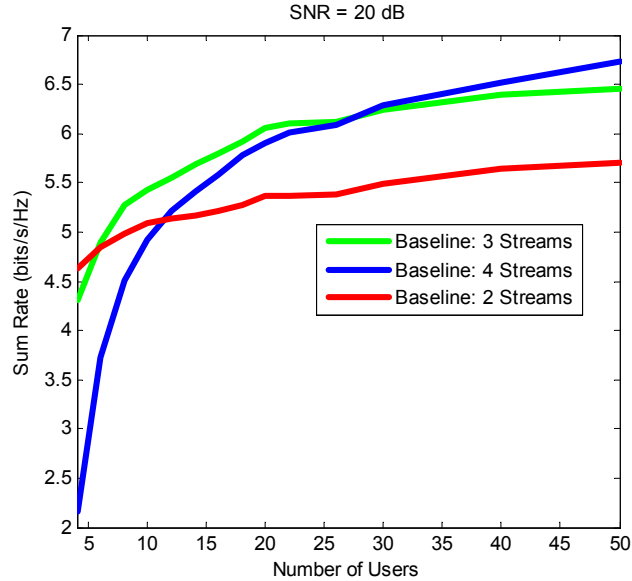, angle=0, width=0.5\textwidth}}
\end{center}
\caption{The effect of the number of streams upon the sum-rate performance for matched filtering (baseline): 19 hexagonal wrap-around-cell layout (${\sf SNR} = 20$ dB).} \label{fig:BLStreams}
\end{figure}

The number of streams is related to the effect of scheduling. We investigate the relationship through simulations.
Fig. \ref{fig:BLStreams} shows the sum-rate performance for the matched filtering (baseline) as a function of $K$.
Note that with an increase in $K$, using more streams gives better performance. This is because for a large value of $K$, an opportunistic scheduler provides good signal separation and power gain, thereby inducing the high {\sf SINR} regime where multiplexing gain affects the performance more significantly than beam-forming gain does. Notice that for a practical range of $K$ (around 10), using 3 streams provides the best performance.

\section*{Acknowledgment}
We gratefully acknowledge Alex Grokhov, Naga Bhushan and Wanshi Chen for discussions on the heterogeneous network scenario.

\bibliographystyle{IEEEtran}
\bibliography{IA_BIB}

\end{document}